\documentclass[twocolumn,prl,showpacs,preprintnumbers,amsmath,amssymb]{revtex4}

\usepackage{graphicx}
\usepackage{dcolumn}
\usepackage{bm}

\begin{document}
\title{Photonic Clusters}

\author{Jack Ng}
\author{Z. F. Lin}
\altaffiliation{Permanent Address: Department of Physics, Fudan University, Shanghai, China.}
\author{C. T. Chan}
\email[Correspondence Author. ]{E-mail: phchan@ust.hk.}
\author{Ping Sheng}

\affiliation{Department of Physics, Hong Kong University of Science and Technology, Clear Water Bay, Hong Kong, China.}

\begin{abstract}
We show through rigorous calculations that dielectric microspheres can be 
organized by an incident electromagnetic plane wave into stable cluster
configurations, which we call photonic molecules. The long-range optical 
binding force arises from multiple scattering between the spheres. A 
photonic molecule can exhibit a multiplicity of distinct geometries, 
including quasicrystal-like configurations, with exotic dynamics. Linear 
stability analysis and dynamical simulations show that the equilibrium 
configurations can correspond with either stable or a type of quasi-stable 
states exhibiting periodic particle motion in the presence of frictional 
dissipation.
\end{abstract}\date{\today}\maketitle

In an optical field of non-uniform intensity, electromagnetic forces can 
move small particles towards regions of high intensity, and such optical 
gradient forces \cite{Ashkin:1980,Chu:1986,Ashkin:1986} 
have been employed fruitfully to manipulate small particles 
\cite{Chu:1998,Ashkin:2000}. Burns \textit{et. al.} \cite{Burns:1990} have 
demonstrated the existence of another weaker optical binding force between 
two dielectric particles, induced by the multiple scattering. In contrast to 
van der Waals force, the optical force can be either attractive or 
repulsive. While Burns \textit{et. al.} have already demonstrated the existence of such a 
force between two particles, and provided an understanding within a 
simplified dipole picture, we here show through rigorous calculations that 
such kind of forces can organize a collection of particles into stable 
equilibrium configurations that behave like ``molecules''. The photonic 
molecules have scale length corresponds to the light wavelength, and the 
binding comes from multiple-scattering induced long range force that is 
strong enough to overcome van der Waals forces and thermal fluctuations. The 
same number of particles can be stabilised in a large variety of distinct 
shapes and bond lengths, and exhibit a multiplicity of static and drifting 
equilibrium configurations. What is rather amazing is that the clusters can 
have distinct shapes and structures, and have well-defined vibration 
frequencies, and yet the structural order is derived from an incident 
electromagnetic plane wave that has uniform intensity (in contrast to 
gradient force that this derived from a non-uniform field). Since the 
optical forces are non-conservative, the dynamical characteristics are very 
different from ordinary molecules bound by chemical forces; and the 
equilibrium configurations can correspond with either stable or a type of 
quasi-stable states in which the molecule maintains an average shape but 
individual particles exhibits periodic motion in the presence of frictional 
dissipation. Besides theoretical interest, the concept of a photonic 
molecule may offer an alternative way to manipulate ultra-fine particles 
into artificial structures. We also note that light-induced forces in 
periodic photonic crystals has been calculated and discussed by 
Antonoyiannakis and Pendry \cite{Antonoyiannakis:1997}.

Consider a cluster of $N$ identical dielectric spheres illuminated by an 
incident time-harmonic EM field 
$\mathord{\buildrel{\lower3pt\hbox{$\scriptscriptstyle\rightharpoonup$}}\over 
{E}} _{inc} 
(\mathord{\buildrel{\lower3pt\hbox{$\scriptscriptstyle\rightharpoonup$}}\over 
{x}} ,t) = \mbox{ }E_0 \hat {x}\mbox{ }\exp (ikz - i\omega t) + E_0 \hat 
{x}\mbox{ }\exp ( - ikz - i\omega t)$, consisting of two counter-propagating 
$x$-polarized monochromatic beams of the same intensity, with angular frequency 
$\omega $ and $k = \omega / c$. The two beams interfere to form fringes with 
intensity varying as $\sim \cos (kz)$, so that the gradient force along the 
$z$-direction traps the spheres in the $xy$-plane \cite{Either:1}. In order 
to be definitive, in what follows the dielectric spheres have radii $r_s = 
0.414$ $\mu m$, mass density 1050 kg/m$^{3}$, dielectric constant $\mbox{ 
}\varepsilon _r = 2.53$ (polystyrene), and the incident wavelength $\lambda = 
0.52 \mu $m ($kr_s = 5)$. 

The intensity of the incident light is uniform in the plane, hence the 
forces on the spheres can only arise from multiple scattering. The light 
induced force (hence the strength of optical binding) is proportional to the 
intensity of the incident light, set to be 0.01 W/$\mu $m$^{2}$, similar to 
that used in the experiments in Ref. \cite{Burns:1990}. We note that the 
physics of the problem remains qualitatively similar when the relevant 
parameters (sphere size, wavelength etc) are varied to within an order of 
magnitude, and a single plane wave is employed instead of two. In the latter 
all the phenomena are reproduced in the frame co-moving with the cluster 
along the $z$ direction. 

We calculate the time-averaged electromagnetic force responsible for optical 
binding. The multiple scattering approach (MS) \cite{The:1995} is adopted 
to compute the electromagnetic fields, then the time-averaged force on the 
sphere $i$, $ < 
\mathord{\buildrel{\lower3pt\hbox{$\scriptscriptstyle\rightharpoonup$}}\over 
{F}} _i > _t $ is given by a surface integral of the time-averaged Maxwell 
stress tensor $ < 
\mathord{\buildrel{\lower3pt\hbox{$\scriptscriptstyle\leftrightarrow$}}\over 
{T}} > _t $ over the sphere's surface: $ < 
\mathord{\buildrel{\lower3pt\hbox{$\scriptscriptstyle\rightharpoonup$}}\over 
{F}} _i > _t = \oint_{surface\mbox{ }of\mbox{ }sphere\mbox{ }i} { < 
\mathord{\buildrel{\lower3pt\hbox{$\scriptscriptstyle\leftrightarrow$}}\over 
{T}} > } _t \cdot 
d\mathord{\buildrel{\lower3pt\hbox{$\scriptscriptstyle\rightharpoonup$}}\over 
{S}} $ The MS-MST approach is highly accurate (retardation effect is 
included) and efficient, and it is the method of choice for a cluster of 
spheres in an arbitrary arrangement.

Figure 1A shows the force for a two-sphere cluster with the bi-sphere axis 
tilted at 45$^{o}$ to the polarization. The van der Waals force 
\cite{adopt:1991} is included in the calculation. The alternating 
attractive-repulsive nature is a manifestation of the phase of the EM 
fields, and the radial force is seen to be stronger than the transverse 
force. The magnitude of the force is inversely proportional to the spheres 
separation in the far zone, as the scattered radiation field decay inversely 
from the scatterer. This long range force, on the order of a pico-Newton, 
dominates over the short-range attractive van der Waals force at distances 
over a wavelength and oscillates between attraction and repulsion with a 
period equal to the wavelength of the incident radiation. Stable radial 
positions are indicated by arrows in Fig. 1A. They correspond to zero-force 
separations in which an increase in the radial separation would induce an 
attractive restoring force, whereas a reduction would induce a repulsive 
restoring force. 

The magnitude of the optical force depends on the radius of the spheres. It 
can be fitted asymptotically to $\vec {f}(r_s )\cos [kR + \phi (r_s )] / R$, 
where $R$ is the separation between the spheres, assumed to be $ \gg r_s $. 
For the polarization perpendicular to the bi-sphere axis, $\vert \vec 
{f}(r_s )\vert $ is shown in Fig. 1B. When $kr_s \ll 1$, the force is 
proportional to the sixth power of the radius. For \textit{kr}$_{s} \ge 1$ the force 
exhibits complex variations, owing to resonant excitations. 

\begin{figure}[htbp]
\centerline{\includegraphics[width=3.24in,height=2.80in]{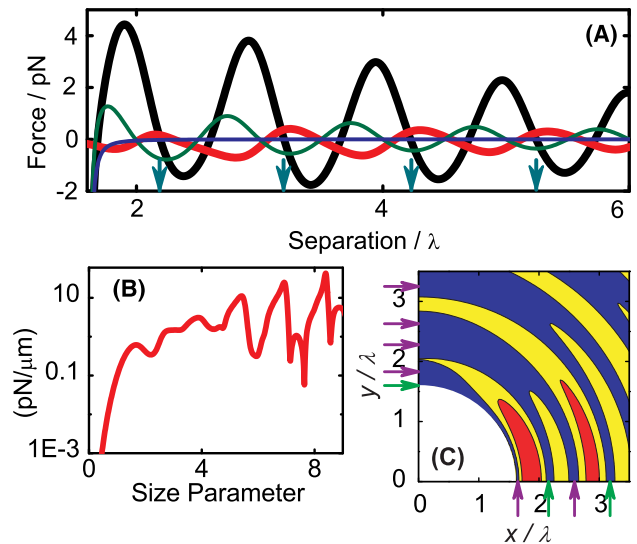}}
\caption{. (A) Forces on two dielectric spheres, plotted as a function of their 
separation. The bi-sphere axis is tilted at 45$^{O }$to the polarization. 
The total radial force (optical plus van der Waals) and transverse force are 
shown by the black and red lines, respectively. Positive radial force means 
repulsion while a positive transverse force means the tendency to align the 
bi-sphere chain along the incident polarization. The green line denotes the 
radial force computed from the dipole approximation with radiative 
correction \cite{Depasse:2001}. Significant difference in magnitude is 
seen. The thin blue line shows the van der Waals force. The arrows indicate 
the positions of stable radial separations. (B): $\vert \vec {f}(r_s )\vert 
$ (see text) versus size parameter (kr$_{s})$. The spheres are aligned 
perpendicular to the polarization of the incident light and located at each 
other's far field zone. (C) Magnitude of the optical force on sphere 2, with 
sphere 1 at the origin. The white region is forbidden (the spheres overlap).
Blue, yellow and red corresponding to weak, medium and strong force respectively.
Zero-force positions are marked by the arrows, with green indicating stable 
positions and purple indicating unstable position.}
\label{fig1}
\end{figure}

In Fig. 1A, it is seen that the stable radial positions correspond to a 
positive aligning force along the transverse direction, therefore it is 
transversely unstable. The zero-force configurations can only be identified 
through a full two dimensional scan of the relevant forces, shown in Fig. 
1C. Multiple zero-force configurations exist on both the $x$ and the $y$ axes. For 
small separations, the force is much stronger when the bi-sphere axis is 
aligned with the polarization, due to the evanescent wave contribution. For 
larger separations, the absence of propagating waves parallel to the 
polarization direction(s) means that the force is stronger when the 
bi-sphere axis is perpendicular to the polarization \cite{general:1}.

Zero force configurations are not necessarily stable. Perturbation 
calculations are required to locate stable equilibrium configurations. For 
general considerations, we assume a damping force proportional to the 
velocity, as in most practical situations. Hence, for a cluster close to a 
zero force configuration denoted by the position vector $\vec {x}_\ast = 
(x_1 ,x_2 , - - - ,x_{2i - 1} ,x_{2i} , - - - )$ \cite{For:1}, where 
$(x_{2i - 1} ,x_{2i} )$ denotes the zero-force coordinate $(x,y)_i $ of the 
$i^{th}$ sphere, the linearized equation of motion is $m\frac{d^2\Delta 
\mathord{\buildrel{\lower3pt\hbox{$\scriptscriptstyle\rightharpoonup$}}\over 
{x}} }{dt^2} \approx 
\mathord{\buildrel{\lower3pt\hbox{$\scriptscriptstyle\leftrightarrow$}}\over 
{K}} \Delta 
\mathord{\buildrel{\lower3pt\hbox{$\scriptscriptstyle\rightharpoonup$}}\over 
{x}} - b\frac{d\Delta 
\mathord{\buildrel{\lower3pt\hbox{$\scriptscriptstyle\rightharpoonup$}}\over 
{x}} }{dt}$, where $m$ is the sphere mass, $\left( {\Delta x_{2i - 1} ,\Delta 
x_{2i} } \right)$ is the displacement vector of the $i^{th}$ sphere from the 
equilibrium, $b$ the damping constant, and 
$(\mathord{\buildrel{\lower3pt\hbox{$\scriptscriptstyle\leftrightarrow$}}\over 
{K}} )_{jk} = \frac{\partial 
(\mathord{\buildrel{\lower3pt\hbox{$\scriptscriptstyle\rightharpoonup$}}\over 
{f}} _{light} )_j }{\partial \Delta x_k }$ is the force constant matrix with 
$\mathord{\buildrel{\lower3pt\hbox{$\scriptscriptstyle\rightharpoonup$}}\over 
{f}} _{light} $ being the light induced force. The eigenvalues of 
$\mathord{\buildrel{\lower3pt\hbox{$\scriptscriptstyle\leftrightarrow$}}\over 
{K}} $ dictate the stability of the cluster. Since a cluster bound by an 
incident light is an open system, the eigenvalues $\lambda _i $ can be 
conjugate pairs of complex numbers \cite{The:1}, as indeed found 
numerically with increasing probability as the number of spheres increases.

Besides the two zero eigenvalues associated with 2D translation, if all the 
other eigenvalues are real and negative, the cluster is stable. The 
appearance of real positive eigenvalue(s) or complex eigenvalues with 
$Re(\lambda _i ) > 0$ denotes instability. The eigenmodes for the complex 
eigenvalues are associated with spiral motions. If the damping coefficient 
$b = 0$, one of the modes spirals inwards upon perturbation, approaching the 
equilibrium position, while the other eigenmode spirals outward. If $b$ is 
finite, then $Re(\lambda _i ) < 0$ indicates the existence of a critical 
damping constant $b_{critical} = \sqrt m \vert Im(\lambda _i )\vert / \sqrt 
{\vert Re(\lambda _i )\vert } $ at which the modes become stable for $b > 
b_{critical} $. We denote this type of complex modes as quasi-stable modes. 
The appearance of one or more pairs of complex eigenvalues with $Re(\lambda 
_i ) < 0$, with all the rest real and negative, denotes the cluster to be 
quasi-stable, i.e., potentially stable in the presence of damping. We note 
that the gradient trapping force in the $z$-direction is strong enough so that 
the system is stable if we consider the cluster as a 3D object.

For a multi-sphere cluster, the optical force can bind the spheres into a 
multiplicity of distinct geometries. In Fig. 2, the stable configurations 
are shown in panels (A)-(E), and quasi-stable configurations are shown in 
panels (F) and (G). For the particular stable configuration shown in Fig. 
2A, the radial oscillation frequency is 540 kHz (\textit{kT} of energy, where $T$ denotes 
room temperature, implies a fluctuation amplitude $\sim $14 nm), whereas the 
rotational oscillation frequency is 51 kHz (\textit{kT} of energy implies a fluctuation 
amplitude $\sim $0.09 radian).

Comparison with molecules bound by chemical forces yields some interesting 
differences. First, atomic clusters can have multiple stable isomers, but 
the bond length always has a narrow distribution. For photonic molecules, a 
fixed number of spheres can be bounded into a great number of stable 
configurations, each of different bound length and shape. Second, chemically 
bound molecules can rotate freely in space while photonic molecules, owing 
to the vector nature of light, are pinned in the two dimensional plane with 
a fixed orientation (relative to the light polarization). Third, for 
clusters without inversion symmetry, photon scatterings are biased in one 
direction and hence the cluster absorbs light momentum, leading to 
``drifting'' equilibrium states in which the cluster moves as a stable 
entity in the \textit{xy}-plane. Thus configurations show in Fig. 2B and Fig. 2E are in 
drifting equilibrium while all the others are in static equilibrium.

\begin{figure}[htbp]
\centerline{\includegraphics[width=3.47in,height=1.37in]{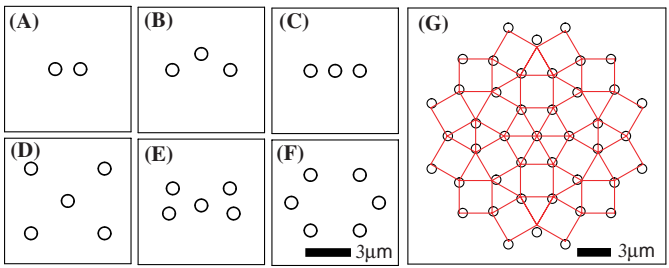}}
\caption{. Examples of photonic molecules. Configurations (B) and (E) are in drifting 
equilibrium (see text), and the others are in static equilibrium. All 
eigenmodes of configurations (A)-(E) are stable, and (F)-(G) have either 
stable or quasi-stable modes. Size of the spheres is drawn to scale with 
respect to the wavelength. Polarization of the incident light beam is in the 
horizontal direction. (G) shows a 43-sphere photonic molecule. The red lines 
are drawn to unveil the 12-fold symmetric, square-triangle quasicrystal 
tiling.}
\label{fig2}
\end{figure}

To study the dynamics of the quasi-stable photonic molecules, it is 
necessary to go beyond the linear stability analysis \cite{The:1994}. We 
integrate the equations of motion using an adaptive time-step 
Runge-Kutta-Verner algorithm. The procedures are similar to molecular 
dynamics simulations, except that the inter-particle forces are now the 
light-induced forces. To illustrate the rich phenomena in dynamics, we study 
a six-sphere photonic molecule shown in Fig. 2F, which has all real negative 
eigenvalues except for one complex conjugate pair. We ignore van der Waals 
forces in the following calculations since the spheres are about 3 microns 
apart. At that distance the optical binding force is about 3000 times 
stronger than the van der Waals force. Figure 3 shows the cluster dynamics 
at different levels of damping. For $b > b_{critical} $, each sphere 
exhibits damped oscillations that settle into the stable zero-force position 
(Fig 3D). For $b$ slightly less than $b_{critical}$, the zero-force position 
turns unstable and an attracting elliptic periodic orbit is formed (Fig 3C, 
period=15 $\mu $s) for each sphere in the immediate neighbourhood of the 
zero force position. This periodic dynamics arise because the repulsive part 
of the force becomes less than that predicted by the linear stability 
analysis as the sphere spirals outward, allowing an orbit to be established. 
Here the damping dissipation is counterbalanced by the incident light 
energy. Hence from the dynamics perspective $b_{critical}$ is a supercritical 
Hopf bifurcation point \cite{The:1994}. A further decrease in $b$ 
enlarges and deforms the periodic orbit (Fig 3B, period=32 $\mu $s), and 
additional decrease in damping leads to instability (Fig 3A). 

\begin{figure}[htbp]
\centerline{\includegraphics[width=3.49in,height=3.41in]{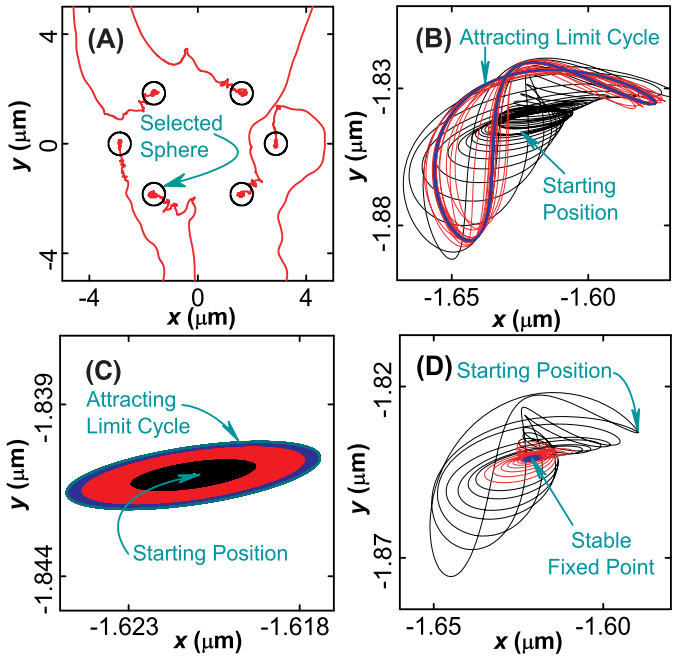}}
\caption{. Complex mode (centre-of-mass) trajectories at different levels of damping 
for the cluster shown in Fig. 2F. (A): Trajectory of the spheres with 
$\mbox{ }b = 0$, the cluster breaks up. (B)-(D) shows the trajectory of the 
selected sphere marked in (A). Black, red and blue colours respectively 
delineate the three consecutive time intervals of equal duration, from early 
to late. (B): For $b = 3.71$ pN/ms$^{ - 1}$, in three 0.5 ms intervals. The 
cluster is attracted towards a periodic orbit. (C): For $b = 4.31$ pN/ms$^{ 
- 1 (}b_{critical} = 4.32$ pN/ms$^{ - 1})$, in three 130 ms intervals. The 
sphere is attracted towards an elliptic periodic orbit. (D): For $b = 9.28$ 
pN/ms$^{ - 1}$, in three 0.2 ms intervals. The cluster is damped towards the 
stable zero-force position. The size of the periodic orbit is noted to be 
small compared to the interparticle separations.}
\label{fig3}
\end{figure}

To investigate the thermal stability of the (stable) photonic molecules, we 
move the spheres apart along the path of their lowest frequency eigenmode, 
until the projection of the restoring force changes sign \cite{For:2}. 
For typical photonic molecules the required dissociation energy per sphere 
was evaluated to be hundreds of \textit{kT} ($T$= room temperature), e.g., the cluster in 
Fig. 2A (two spheres at distance=3.16$\lambda )$ has a dissociation energy 
of 110 \textit{kT}. Thus the photonic molecules are expected to be stable against 
thermal fluctuations.

Although the input light field is homogeneous on the \textit{xy}-plane, it is rather 
intriguing that the optical binding force can arrange dielectric spheres 
into rather complex geometries---Fig. 2G shows a 43-sphere cluster with a 
quasicrystal-like geometry (red lines show the exact square-triangle 
quasicrystal tiling). The particular configuration shown has 18 pairs of 
complex eigenvalues with their corresponding $b_{critical}$ between 1.3 and 
17 pN/ms$^{ - 1}$. Thus if $b > 17$ pN/ms$^{ - 1}$, a stable 
quasicrystal-like cluster can be realized. 

In summary, we showed using a rigorous multiple-scattering technique that 
the homogenous light-field of an incident plane wave can organize small 
particles into stable clusters. These clusters have well-defined geometry 
(bond length and angle) and well-defined vibration frequencies, and they are 
bound by light, and hence we call them photonic molecules. Both the static 
and dynamic properties are rather interesting, which is a manifestation that 
we have an open system, in contrast to chemical molecules are that bound by 
conservative forces. 

Support by RGC Hong Kong through HKUST6138/00P is gratefully acknowledged. ZF Lin is supported by CNKBRSF and China NNSF 10321003 and 10474014. We thank Dr. L. M. Li for help in the molecular dynamics code.


\begin{thebibliography}{13}
\bibitem{For:1} For the drifting equilibriums introduced in this work, the same perturbation analysis can still be applied except now $\Delta \mathord{\buildrel{\lower3pt\hbox{$\scriptscriptstyle\rightharpoonup$}}\over {x}} $ denotes the displacement from the drifting equilibrium, i.e., \newline
$\Delta \mathord{\buildrel{\lower3pt\hbox{$\scriptscriptstyle\rightharpoonup$}}\over {x}} = \mathord{\buildrel{\lower3pt\hbox{$\scriptscriptstyle\rightharpoonup$}}\over {x}} (t) - \mathord{\buildrel{\lower3pt\hbox{$\scriptscriptstyle\rightharpoonup$}}\over {x}} _0 (t)\quad \mbox{and}\quad \mathord{\buildrel{\lower3pt\hbox{$\scriptscriptstyle\rightharpoonup$}}\over {x}} _0 (t) = \mathord{\buildrel{\lower3pt\hbox{$\scriptscriptstyle\rightharpoonup$}}\over {x}} _\ast + \left\{ {\begin{array}{l}
 \mathord{\buildrel{\lower3pt\hbox{$\scriptscriptstyle\rightharpoonup$}}\over {f}} _{light} (\mathord{\buildrel{\lower3pt\hbox{$\scriptscriptstyle\rightharpoonup$}}\over {x}} _\ast )t / b\mbox{ }if\mbox{ }b \ne 0 \\ 
 \mathord{\buildrel{\lower3pt\hbox{$\scriptscriptstyle\rightharpoonup$}}\over {f}} _{light} (\mathord{\buildrel{\lower3pt\hbox{$\scriptscriptstyle\rightharpoonup$}}\over {x}} _\ast )t^2 / 2m\mbox{ }if\mbox{ }b = 0 \\ 
 \end{array}} \right.$.
\bibitem{The:1} The force is not conservative, so it cannot be expressed as the gradient of a potential. Thus the force constant matrix is not necessarily symmetric, and conjugate pairs of complex eigenvalues are possible.
\bibitem{For:2} For an open system, the work done is path dependent. Here we give a plausible estimate to the depth of the equilibrium well.
\end{thebibliography}
\end{document}